\documentclass[preprint]{elsarticle}

\usepackage{lineno,hyperref}
\usepackage{graphicx}
\usepackage{amsmath}
\usepackage{amssymb}
\usepackage{array}
\usepackage[colorinlistoftodos]{todonotes}

\usepackage[utf8]{inputenc}

\newcommand{\D}[0]{\mathrm{d}}

\modulolinenumbers[0]

\makeatletter
\def\ps@pprintTitle{%
 \let\@oddhead\@empty
 \let\@evenhead\@empty
 \def\@oddfoot{\centerline{\thepage}}%
 \let\@evenfoot\@oddfoot}
\makeatother

\begin{document}
\begin{frontmatter}
\title{ Disk wrinkling under gravity}

\author{\large Gwenn Boedec}
\author{\large Julien Deschamps}
\address{\large  Aix Marseille University, CNRS, Centrale Marseille, IRPHE, Marseille, France}



\begin{abstract}
We study the deflection by gravity of a circular elastic disk deposited on a rigid support. The axisymmetric deflection induces compressive orthoradial stresses which leads to a wrinkling instability above a critical threshold of the dimensionless gravity force. We study this instability by a combination of experiments, numerical simulations and analytical tools, with a particular focus on the role of geometry. We show that aspect ratio is a crucial parameter that controls both the threshold of instability and the most unstable mode. The influence of this parameter on the threshold can be catched by introducing a new nondimensionalization of the transverse load.\end{abstract}


\end{frontmatter}


\section{Introduction}


Depositing a circular napkin on a circular table with a smaller radius, one may observe that the outer boundary undulates, with a wavelength depending on the ratio of the napkin radius and the table radius. This is a manifestation of an elastic instability called wrinkling, where the system releases some stretching energy by developing out-of-plane undulations.
Many different systems may develop wrinkles, like stretched \cite{Cerda_2002} or sheared \cite{Wong_2006} rectangular elastic plates, tensed circular elastic films under indentation \cite{Huang_2007,Holmes_2010,Paulsen_2016}, or deposited on a droplet \cite{King_2012}. One prototypical situation  to study wrinkling is the Lam\'e setup, where an annulus of elastic material is submitted to radial tension on its edges \cite{Davidovitch_2011}. Depending on the differential tension, a zone of compressive hoop stress may develop, which leads ultimately to the formation of wrinkles. Role of geometry is intricate : theoretical analysis \cite{Coman_2007} shows that aspect ratio has an importance on the wavenumber selection, and that pockets of instability overlap, which may lead to multistable systems. While the Lam\'e setup is well-defined theoretically, its experimental realization is trickier since it requires to induce a differential stresses at the edges while still keeping the sheet planar (before wrinkling) : this has been recently realized by using surface tension variations with surfactants \cite{Pineirua_2013}. 

In other situations, the sheet is generally deformed out of plane, and this deformation might also generate compressive hoop stress : in a setup such as the napkin on the table, the gravity-induced deflection of the sheet is in fact the driving mechanism for the development of a region under compression. A similar situation may be encountered in the context of lightweight deployable space structures, like solar sails or solar power satellites \cite{Delapierre_2018,Sader_2019}. In this case, the weight is replaced by the load of the solar light  so that maintaining an (almost) flat shape is important for efficiency, and may be attained by using centrifugal forces. 

In this paper, we study gravity-induced wrinkling of the overhang part of an elastic disk deposited on a rigid circular support of a smaller size. A theoretical study by \cite{Cerda_2004} has shown that in the limit of a punctual support, there is a bifurcation between a one-folded shape similar to a d-cone, and a two-folded shape. This study was mostly concerned with asymptotic behavior in the fully non-linear regime (large deflection). On the other hand, \cite{Chen_2010} study theoretically (numerically) the deformation under gravity of a circular plate supported by an inner ring, in the small to moderate deflection regime. They show that first an axisymmetric solution exists (no wrinkles), and that, for further loading, this solution destabilizes, with a critical mode number and a critical load depending on the aspect ratio. \cite{Chen_2011} then extended the theoretical study to include the effect of rotation and conducted some experiments showing the coexistence of $\cos 2 \theta$ and $\cos 3 \theta$ deformations. Recently, 
\cite{Delapierre_2018} conducted a detailed analysis of spinning transversely loaded membranes,  coupling experiments and numerical analysis with a buckling analysis of F\"oppl-von Karman equations to study the effect of rotation on the wrinkling patterns. In the same manner, \cite{Sader_2019} considered the impact of the spin-up of the disk on the buckled modes. Despite the strong effect of the aspect ratio on the critical load and on the observed wavenumber, both \cite{Chen_2011} and \cite{Delapierre_2018} conducted experiments with a unique fixed aspect ratio (0.1 for \cite{Delapierre_2018} and 0.3 for \cite{Chen_2011}), focusing mostly on the importance of rotation : No systematic experimental diagram study was conducted. As a first step the present study is concerned with the influence of aspect ratio for a non-rotating disk, deflected only by gravity. We systematically vary the relevant parameters to obtain an experimental phase diagram of the system, and show that the two dimensionless parameters can be collapsed in a single one controlling the transition between wrinkled and unwrinkled state. We rationalize this finding by developing an original asymptotic analysis which predicts analytically the instability criteria. These results are systematically validated by comparison with a buckling analysis close to the threshold and full numerical simulations of thin-shell equations for the nonlinear regime.
\newline

\section{System description}
\begin{figure}
\begin{tabular}{>{\centering\arraybackslash}m{.32\textwidth}>{\centering\arraybackslash}m{.6\textwidth}}
\includegraphics[scale=0.13]{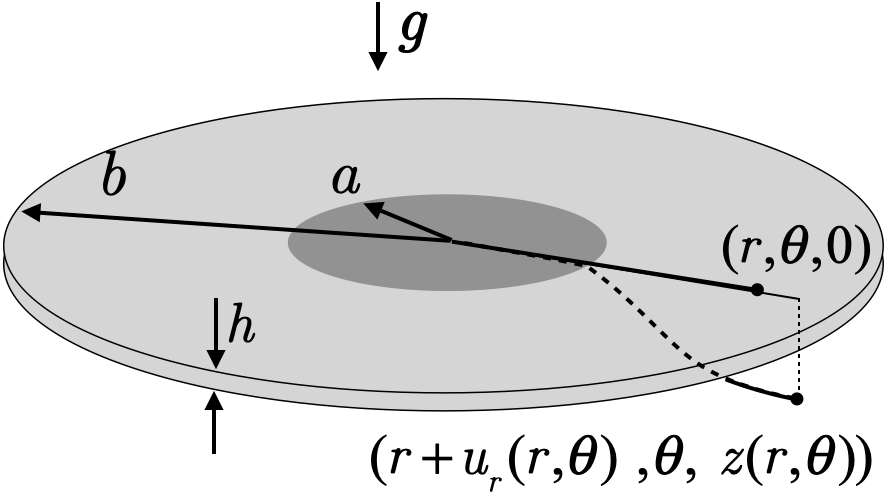}
&
\includegraphics[scale=0.15]{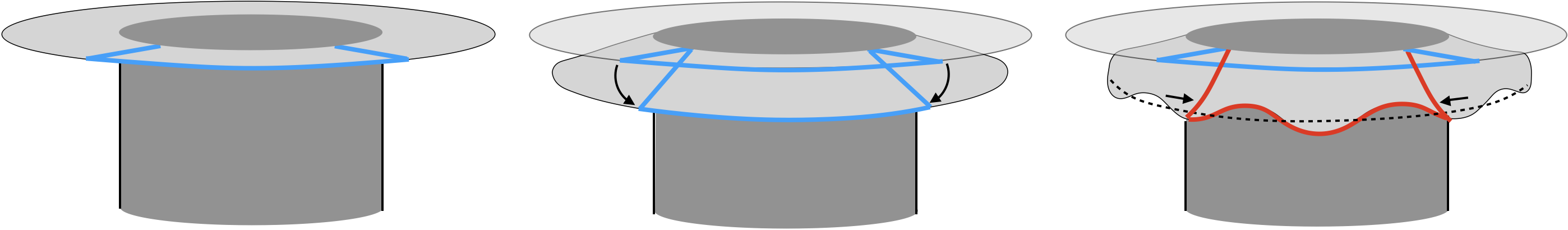}\\
a&b
\end{tabular}
\caption{ a. Definition of the system. b. Mechanism of wrinkling instability : considering an angular sector of the disk (blue lines), the vertical deflection implies a diminution of the length at the outer edge. The induced compressive strain can be released by allowing undulations.}
\label{systemdefinition}
\end{figure}

We consider a disk of radius $b$ made of an elastic material. The thickness of the disk $h$ is such that $h/b \ll 1$, thus the disk may appropriately be described as a thin plate whose non-deformed shape is planar. This disk is deposited on a circular rigid support of radius $a<b$ and we let it hang freely under the action of gravity. Our system is thus equivalent to an annulus of elastic material, whose inner edge is clamped and whose outer edge is free (see figure \ref{systemdefinition} a). While this clamped boundary condition is not strictly enforced experimentally we checked that no detachment of the inner part occurs in the data reported in this paper.

Due to gravity the annulus deflects downwards. For very small deflections, the shape remains axisymmetric as the annulus adopts a conical shape, but as the deflection, i.e. the effect of gravity, increases, the outer edge starts to undulate orthoradially as a wrinkling instability develops on the annulus (see figure \ref{systemdefinition} b). This instability can be simply understood as it is not possible for the outer edge to be deflected without stretching, either radially or orthoradially, the sheet. A part of the resulting stretching energy due to compressive hoop stress can be relaxed by developing out-of-planes undulations at the expense of some bending energy that is energetically favorable since the sheet thickness is small. Following \cite{Chen_2010,Chen_2011,Delapierre_2018}, two dimensionless parameters control this instability. First, a dimensionless measure of gravity $G$ is defined as :
\begin{equation*}
G=\frac{12 (1-\nu^2) \rho g b^4 }{E h^3 }
\end{equation*}
where $\rho$ is the volume mass of the disk, $g$ is the acceleration due to gravity, $E$, $\nu$ are the Young modulus and Poisson coefficient of the material. Second, a geometrical parameter is involved, the aspect ratio of the annulus :
\begin{equation*}
 \alpha = \dfrac{a}{b}
 \end{equation*}
 

\section{Methods}
\subsection{Experiments}

We use disks cut into elastic sheets made of four different materials : PDMS, Silicone, Latex and Mylar. Latex disks are cut from commercial dental dam  and Mylar disks are obtained from standard stencil sheets.
PDMS and Silicone sheets are formed via the same process : we mix two liquid components, the polymer base and the curing agent. (Silgard 184 purchased from Merck for PDMS and RTV 181 from Esprit Composite for Silicone). We put the mixture into a vacuum chamber to remove air bubbles and then mold the mixture between two planar parallel plates made of aluminium with a controlled spacing. We then let the mixture cure for 24h-72h at room temperature. The elastic sheet is then gently detached from the mold. Whatever the material, the thickness of the disk is measured in different locations with a Palmer micrometer and averaged. The Young modulus is measured with a homemade tensile test machine. For PDMS, different proportion of the mixture allows to change the Young modulus by an order of magnitude. For each material, diameters, thicknesses and Young moduli ranges are reported in table \ref{tab_material_used}. Once the disk is characterized, it is gently deposited on cylindrical support of radius $a$.


\begin{table}[!h]
    \centering
    \begin{tabular}{c|c|c|c}
         Material & $E$ (MPa) & $h$ ($\mu$m) & $b$ (mm)\\
         \hline 
         PDMS & \textbf{0.2} ($\pm$0.02)-\textbf{2.3} ($\pm$0.1)& \textbf{220} ($\pm$8)-\textbf{1800} ($\pm$60) & \textbf{15}-\textbf{120} \ ($\pm$1)\\
         \hline
         Silicone & \textbf{0.5}-\textbf{1} ($\pm$ 0.1) & \textbf{150} ($\pm$6)-\textbf{3200} ($\pm$50) & \textbf{25}-\textbf{135} \ ($\pm$1)\\
         \hline
         Latex & \textbf{2.6} ($\pm$0.1) & \textbf{150}-\textbf{230} ($\pm$20) & \textbf{15}-\textbf{70} \ ($\pm$1) \\
         \hline 
         Mylar & \textbf{4000} ($\pm$200) & \textbf{125} ($\pm$20) & \textbf{55}-\textbf{110} \ ($\pm$1)
    \end{tabular}
    \caption{Range of parameters for each material. Typical uncertainties are indicated in parenthesis}
    \label{tab_material_used}
\end{table}

\subsection{Numerics}

We simulate our setup using an in-house code which solves thin plate equations based on an isogeometric framework. The code is built upon a method using subdivision elements \cite{Cirak_2000,Cirak_2001}, with the improvement of \cite{Green_2004} for dealing with boundary conditions: we use a clamped boundary for the inner edge and a free boundary condition for the outer edge. The plate kinematics are based on Kirchoff-Love theory, assuming that in the deformed configuration, the plate director remains normal to the mid surface. Within this framework, plate deformation is completely given in terms of membrane strains and bending strains defined on the mid surface. Note that we do not restrict the description to linearized kinematics and resolve the non-linear problem with a modified Newton-Raphson solver. For the material behavior, a generalized Hooke law is chosen, meaning that we have a linear elastic material behavior, but with all geometrical nonlinearities. The plate is discretized using Catmull-Clark elements.

\subsection{Stability analysis}
\label{sec_stabilityAnalysis}

We use the method of \cite{Delapierre_2018} to compute the critical load. We briefly summarize the main ideas here, and refer to their work for a complete description of the method. The principle is as follows : complete F\"oppl-von Karman equations \ref{eq:FvK1}-\ref{eq:FvK2} are solved via a boundary value problem solver (e.g. the \texttt{solve\_bvp} function in Python) to first obtain a stationary axisymmetric solution for a given couple $(G,\alpha)$: 

\begin{eqnarray}
\begin{split}
    D\nabla^4 z=&\partial_{rr}z\left(\frac{1}{r}\partial_r\phi+\frac{1}{r^2}\partial_{\theta\theta}\phi\right)+\partial_{rr}\phi\left(\frac{1}{r}\partial_rz+\frac{1}{r^2}\partial_{\theta\theta}z\right)\\
    &-2\left(\frac{1}{r}\partial_{r\theta}\phi-\frac{1}{r^2}\partial_{\theta}\phi\right)\left(\frac{1}{r}\partial_{r\theta}z-\frac{1}{r^2}\partial_{\theta}z\right)+\rho gh
    \label{eq:FvK1}
    \end{split}\\
    \nabla^4 \phi=Eh\left[\left(\frac{1}{r}\partial_{r\theta}z-\frac{1}{r^2}\partial_{\theta}z\right)^2-\partial_{rr}z\left(\frac{1}{r}\partial_{r}z+\frac{1}{r^2}\partial_{\theta\theta}z\right)\right]
    \label{eq:FvK2}
\end{eqnarray}

 where $z(r,\theta)$ is the vertical deflection of the deformed disk (figure \ref{systemdefinition} a) and $\phi(r,\theta)$ is the Airy function related to the stress components:
\begin{align*}
    \sigma_{rr}=\frac{1}{h}\left(\frac{1}{r}\partial_r\phi+\frac{1}{r^2}\partial_{\theta\theta}\phi\right)
    \ \ \ \ \ 
    \sigma_{\theta\theta}=\frac{1}{h}\partial_{rr}\phi \ \ \ \ \ 
    \sigma_{r\theta}=-\frac{1}{h}\partial_{r}\left(\frac{1}{r}\partial_\theta\phi \right)
\end{align*}

The solution satisfies clamped boundary conditions (no displacement, no rotation) at the inner edge and free boundary (no stresses, no moments) at the outer edge. Then, the stability of this axisymmetric solution with respect to buckling is computed by introducing a perturbation with an assumed form $z(r,\theta)=Z(r)\exp(i n \theta)$. 
After linearization of the F\"oppl-von Karman axisymmetric equations, the resulting eigenvalue problem is solved, using again a boundary value solver. 
Depending on the sign of the eigenvalue, a small perturbation would either be damped (stable) or amplified (unstable). By coupling this with a root-finding algorithm, one can find for a given $\alpha$ and mode number $n$ the critical load $G^\star_{crit}$ for which the eigenvalue sign changes.





\subsection{Scaling of instability}
\label{sec_stabilityAnalysis}
To gain some insight in the behavior of the system, we develop a simplified theoretical analysis based on different scalings of the energies. 


We assume the position of a point initially at $(r,\theta,0)$ on the flat disk can be described as 
$(r +u_r,\theta, z(r,\theta))$
(figure \ref{systemdefinition} a) with $u_r$ the radial displacement and  $z(r,\theta) = w(r) (1+A \cos(n\theta))$ the deflection.
With this assumption, and keeping the leading non-linearities due to geometry (only quadratic terms in the deflection, not in displacement), we express strains and curvatures to dominant order:

\begin{equation*}
    \begin{split}
        \epsilon_{rr} &\sim \partial_r u_r + \frac{1}{2}(\partial_r z)^2 \\
        \epsilon_{\theta\theta} &\sim \frac{u_r}{r} + \frac{1}{2r^2}(\partial_\theta z)^2 \\
        \kappa_{rr} &\sim \partial_{rr} z\\
        \kappa_{\theta\theta} &\sim \frac{\partial_r z}{r} +\frac{\partial_{\theta\theta} z}{r^2}\\
    \end{split}
\end{equation*}

There are three energies densities involved in the problem :
\begin{itemize}
    \item stretching energy  $$\mathcal{E}^{s}   \sim \frac{E h }{(1-\nu^2)}\int_{S} \left[\epsilon_{rr}^2+\epsilon_{\theta\theta}^2 + 2 \nu \epsilon_{rr}\epsilon_{\theta\theta}\right] \D S $$ 
    \item bending energy
     $$ \mathcal{E}^{b} \sim \frac{E h^3 }{12(1-\nu^2)}\int_{S} \left[\kappa_{rr}^2+\kappa_{\theta\theta}^2 + 2 \nu \kappa_{rr}\kappa_{\theta\theta}\right] \D S$$
    \item gravity energy
    $$\mathcal{E}^{g} \sim  \int_S{h \rho g z}\D S$$
\end{itemize}

For a given set of parameter, the solution to the problem is given by a minimization of the total energy of the system, which results in F\"oppl-von Karman equations. However, as these equations are notoriously difficult to solve in a general case, we resort to scaling analysis under simplifying assumptions to try to gain some insight in the behavior of the system.

Assume a planar ring of elastic material of inner radius $a$ i.e. $\alpha b$ and outer radius $b$. We denote the typical deflection at the outer boundary by $\delta \ (=w(b))$, the typical size of radial displacement by $U$ and the width of the annulus by $l=b(1-\alpha)$. We neglect the Poisson coefficient effect ($\nu=0$). With these variables, and using a bar over dimensionless quantities, we estimate for instance:
$$\epsilon_{rr}\sim \frac{U}{l} \partial_{\overline{r}}\overline{u} + \frac{1}{2} \left(\frac{\delta}{l}\right)^2 \left(\partial_{\overline{r}} \overline{z}\right)^2$$
where the scaled quantities $\partial_{\overline{r}}\overline{u}$ or $\partial_{\overline{r}} \overline{z}$ are expected to be of order one if the scaling is correctly chosen. For the sake of clarity of the presentation, we will omit the scaled terms in the following, and keep only the scalings. Thus, the previous equation reads :
$$\epsilon_{rr}\sim \frac{U}{l}  + \frac{1}{2} \left(\frac{\delta}{l}\right)^2 $$
Likewise, the orthoradial stretching is estimated as
$$\epsilon_{\theta\theta} \sim \frac{U}{b}+\frac{1}{2b^2} \delta^2 A^2 n ^2 \sin(n \theta)^2 $$
Note that there is a difference in scaling of $\partial_r u \sim \frac{U}{l} \partial_{\overline{r}}  \overline{u}$ and $\frac{u}{r} \sim \frac{U}{b} \frac{\overline{u}}{\overline{r}}$ because for derivation the relevant length scale is the width of the annulus and not its radius.

With these two scalings for the stretching we can estimate for instance the unwrinkled ($A=0$) stretching energy as 
\begin{equation}
\mathcal{E}^{s}\sim Eh \int_{S}[\epsilon_{rr}^2+\epsilon_{\theta\theta}^2]\D S \sim SEh\left[\left(\frac{U}{l}  + \frac{1}{2} \frac{\delta^2}{l^2}\right)^2+\left(\frac{U}{b}\right)^2\right]
\label{eq_estretch0}
\end{equation}
where $S$ is the surface of the annulus.
We proceed to analyse the system in the two asymptotic limits of $\alpha \rightarrow 0$ (point like support) and $\alpha \rightarrow 1$ (quasi 2D system). Details of calculation are provided in \ref{sec_alpha0} and \ref{sec_alpha1}.

\section{Results}

\begin{figure}[h!]
\centering
\begin{tabular}{c}
    \includegraphics[scale=0.23]{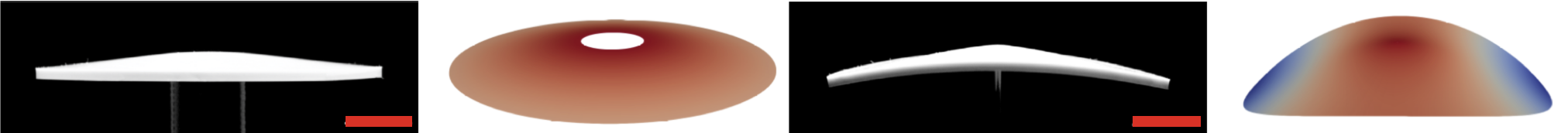}\\
    a\\
    \\
    \includegraphics[scale=0.53]{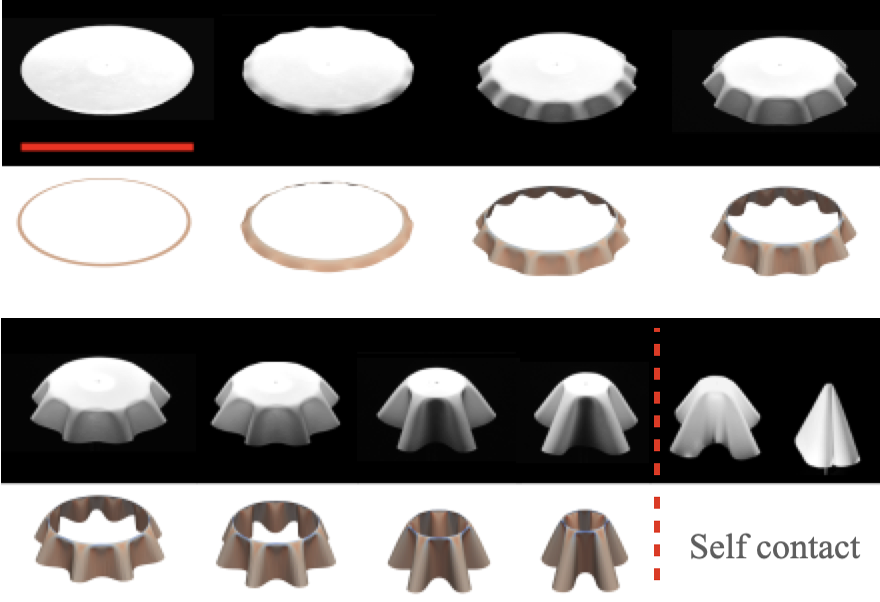}\\
    b\\
    \end{tabular}
    \caption{Comparison of experimental and numerical shapes of deformed disk under gravity. a. Axisymmetric mode ($\alpha$=0.21, $G=65$) and mode 2 ($\alpha$=0.02, $G=65$) for silicone disk. Scale bar is 20 mm b. Comparison of the evolution of wrinkling pattern with a decreasing aspect ratio in experiments (upper row) and in numerical simulations (lower row). From left to right (mode number is indicated in parenthesis) : $\alpha = 0.95 (\mathrm{stable}); 0.88 (n=16); 0.75 (13); 0.66 (11); 0.58 (9); 0.50 (7); 0.33 (6);0.25 (5); 0.17; 0.008$. $G=867 000$. Scale bar is 120 mm.}
    \label{fig_photosXP_num}
\end{figure}

For small deflection of the outer edge, we observe experimentally that the plate adopts an axisymmetric conical shape (figure \ref{fig_photosXP_num} a). 
For a given plate, the maximal deflection increases when the aspect ratio decreases as shown in figure \ref{fig_photosXP_num} b. For a fixed aspect ratio, increasing the dimensionless gravity $G$ also increases the maximal deflection. We observe that when the deflection reaches a critical value, the outer edge develops an undulated shape, revealing a wrinkling instability.

\begin{figure}[h!]
\centering
    \includegraphics[scale=0.25]{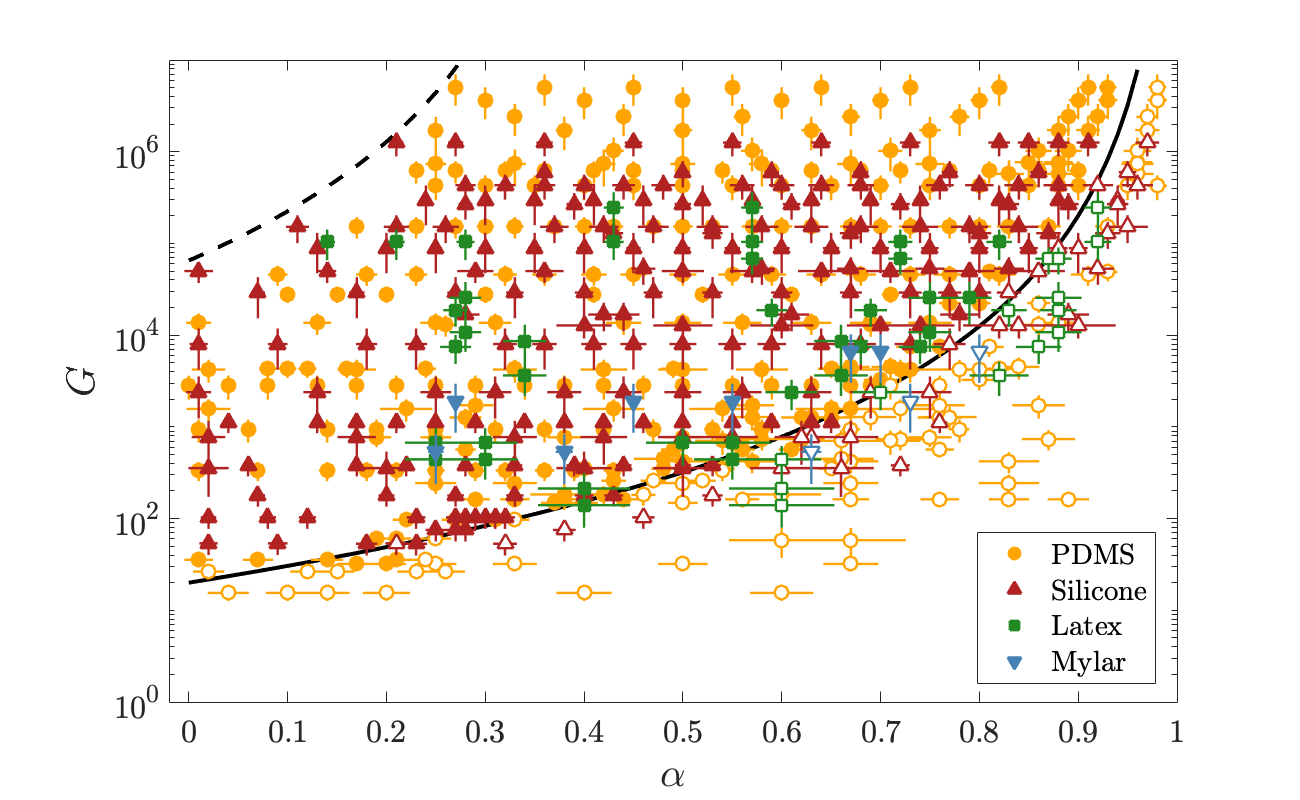}
     \caption{Experimental stability boundary : filled symbols are wrinkled states, open symbols are axisymetric ones. Colors show the different materials used. Dashed line is the limit of self contact mode. Solid line is $G^* = \mathrm{cst}$.}
    
    \label{fig:fig_xp_stabilityDiagramGalpha}
\end{figure}

\begin{figure}[h!]
\centering
    \includegraphics[scale=0.25]{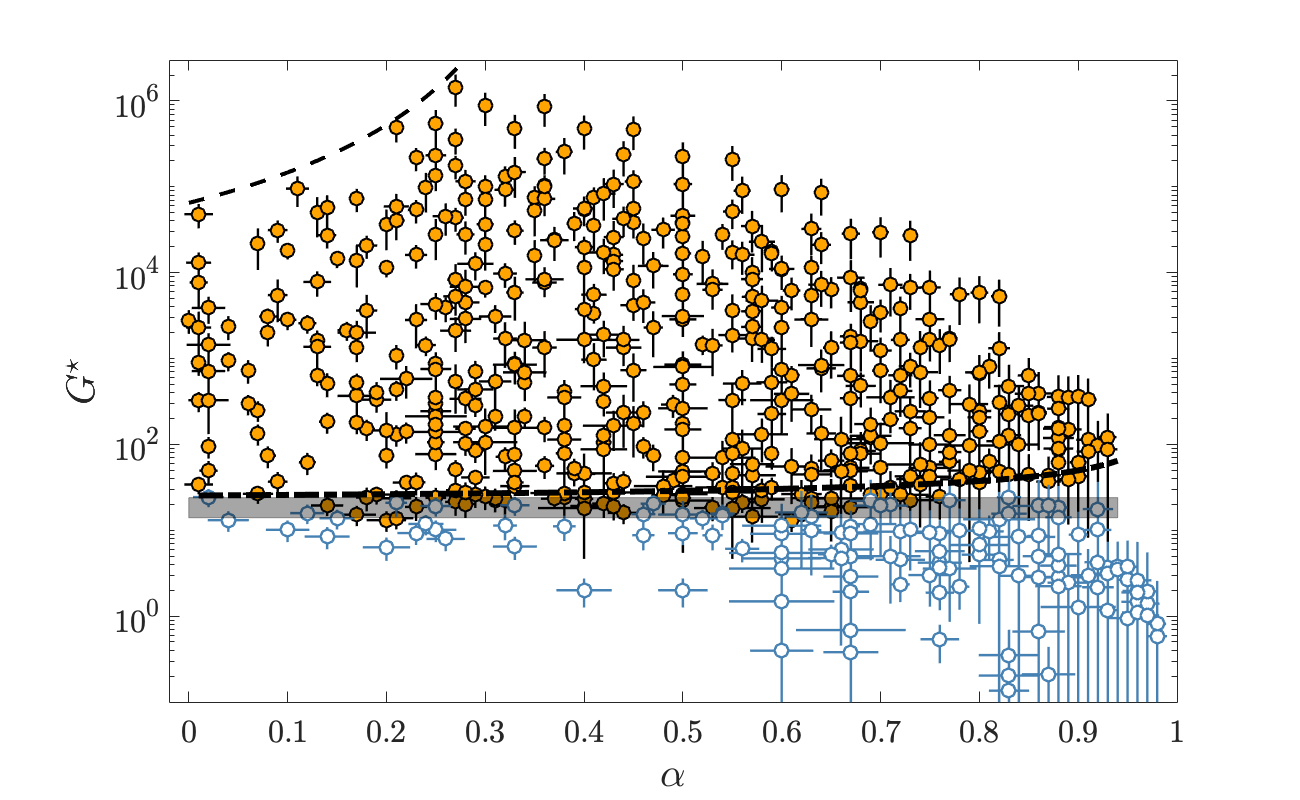}
    
     \caption{Experimental stability boundary : filled symbols are wrinkled states, open symbols are axisymetric ones. Grey solid line is $G^* = 20\pm 5$. Dash-dotted line is the envelope of the linear stability analysis determined in figure \ref{fig_boundariesNumThXP} }

    \label{fig:fig_xp_stabilityDiagramGstaralpha}
\end{figure}

We plot in figure \ref{fig:fig_xp_stabilityDiagramGalpha} all experimental points classified in three categories: self contact, axisymmetric and wrinkled. We focus particularly on the last two and observe that there is a well-defined boundary between these two regions. This frontier appears even clearer as we plot in figure \ref{fig:fig_xp_stabilityDiagramGstaralpha} the whole set of experimental data as $G^\star(\alpha)$ where $G^\star=G(1-\alpha)^4$. 
The key idea behind this scaling is that compressive stress is induced by the deflection of the outer edge, and this deflection is controlled by bending of the width of the plate, not by the bending of the whole plate : this new insight leads to consider the characteristic length scale appearing in the dimensionless parameter to be $(b-a) = (1-\alpha)b$ instead of $b$. Using this new dimensionless parameter to determine the stability boundary shows that the criteria $G^\star = \mathrm{cst}$ is relevant over the whole range of aspect ratio $\alpha$. In figure \ref{fig:fig_xp_stabilityDiagramGstaralpha} the frontier can be fitted as $G^\star\approx20$. 

\begin{figure}[h!]
    \centering
    \includegraphics[scale=0.25]{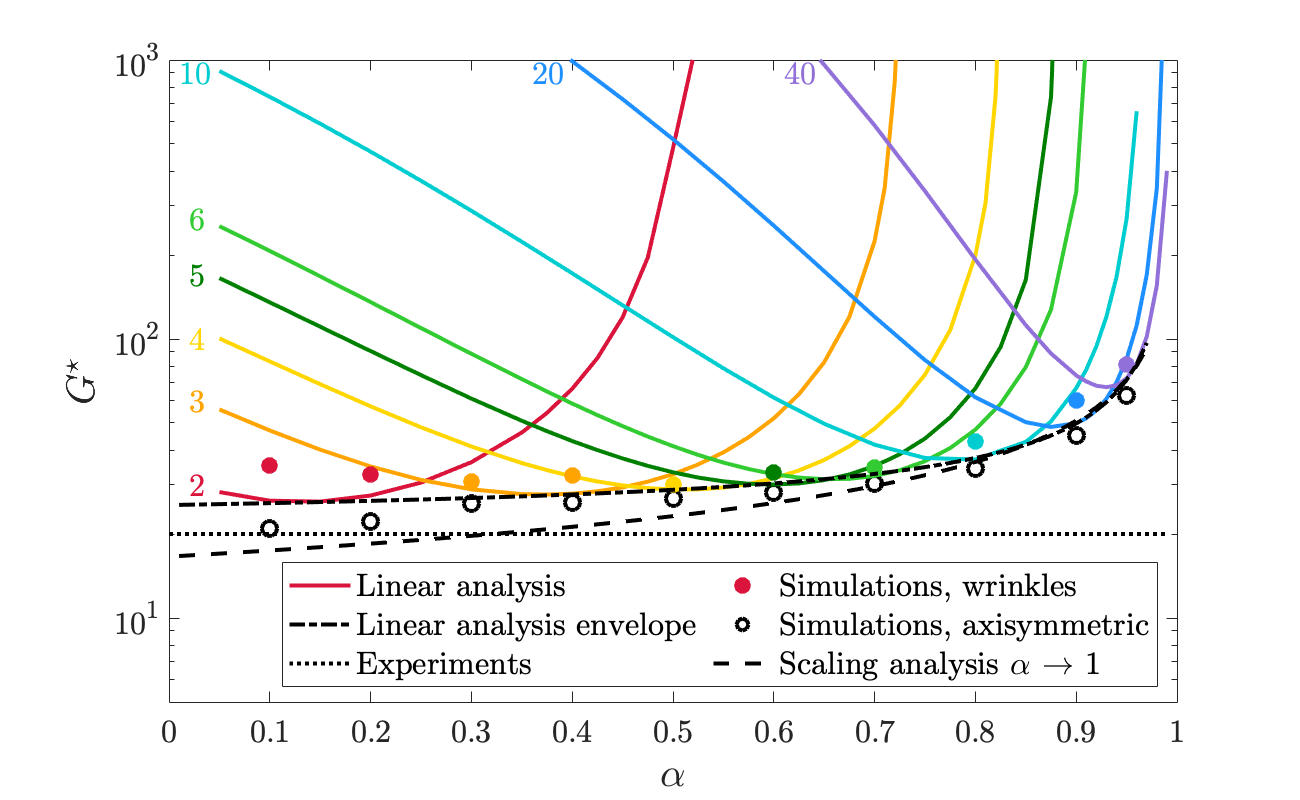}
    \caption{Stability boundaries of different modes computed by a linearized buckling analysis shows an excellent agreement with full numerical solutions of thin shell equations. The envelope of the boundaries (Dash-dotted line is $G^\star=20.8+4.6(1-\alpha)^{-0.8} $) is however, slightly higher than the experimental observation (Dotted line as $G^\star=20$.) The dashed line is the scaling limit $\alpha \rightarrow 1$ :  $G^\star \sim(1-\alpha)^{-1/2}$ (\ref{sec_alpha1}). the fit is performed over the linear stability envelope for $\alpha>0.8$ and gives $G^\star=0.7+15.9(1-\alpha)^{-1/2} $.  }
    
    
    \label{fig_boundariesNumThXP}
\end{figure}

We compare experimental results with predictions coming from linear stability analysis and complete numerical simulations of thin shell equations, as shown in figure \ref{fig_boundariesNumThXP}. Linear stability analysis shows that, for a given mode number $n$, the neutral curve (zero growth rate) $G^\star_{crit}(\alpha)$ is weakly decreasing for small aspect ratios, reaches a minimum and then increases rapidly. Close to threshold, full numerical simulations are in excellent agreement with linear stability analysis, and also show that $G^\star=\mathrm{cst}$ is a relevant criteria for almost all the range of aspect ratio. However both predict a slightly higher threshold ($G^\star\approx 25$).
In the limit $\alpha\rightarrow 1$ we notice a weak increase of the threshold. 
This result is in good agreement with the scaling analysis developed in \ref{sec_alpha1}. 
In the limit $\alpha\rightarrow 0$ the asymptotic analysis presented in \ref{sec_alpha0} shows that the wrinkling instability develops when the outer edge deflection reaches a critical value. In the bending regime, this deflection scales as $G^\star$, which gives the scaling $G^\star \approx cst$. 
In the limit $\alpha\rightarrow 1$ it is shown that $G^\star$ slightly diverges as $(1-\alpha)^{-1/2}$. While these results come from asymptotic analysis and are strictly speaking only valid in their respective limit, we notice that they hold beyond their region of derivation. Note also that the difference between the two scalings is weak: turning back to non-dimensionalization, the $\alpha\rightarrow 0$ limit indicates that one should rescale the length with $(1-\alpha)^4$ while the $\alpha\rightarrow 1$ limit indicates that one should rescale the length with $(1-\alpha)^{4.5}$. This is an effect too much subtle to be of practical impact or even experimentally accessible. Thus for simplicity we propose a criterion $G^\star = \mathrm{cst}$ over the full range of $\alpha$ to be the most adequate. 



\begin{figure}[!h]
\begin{center}
\includegraphics[scale=0.25]{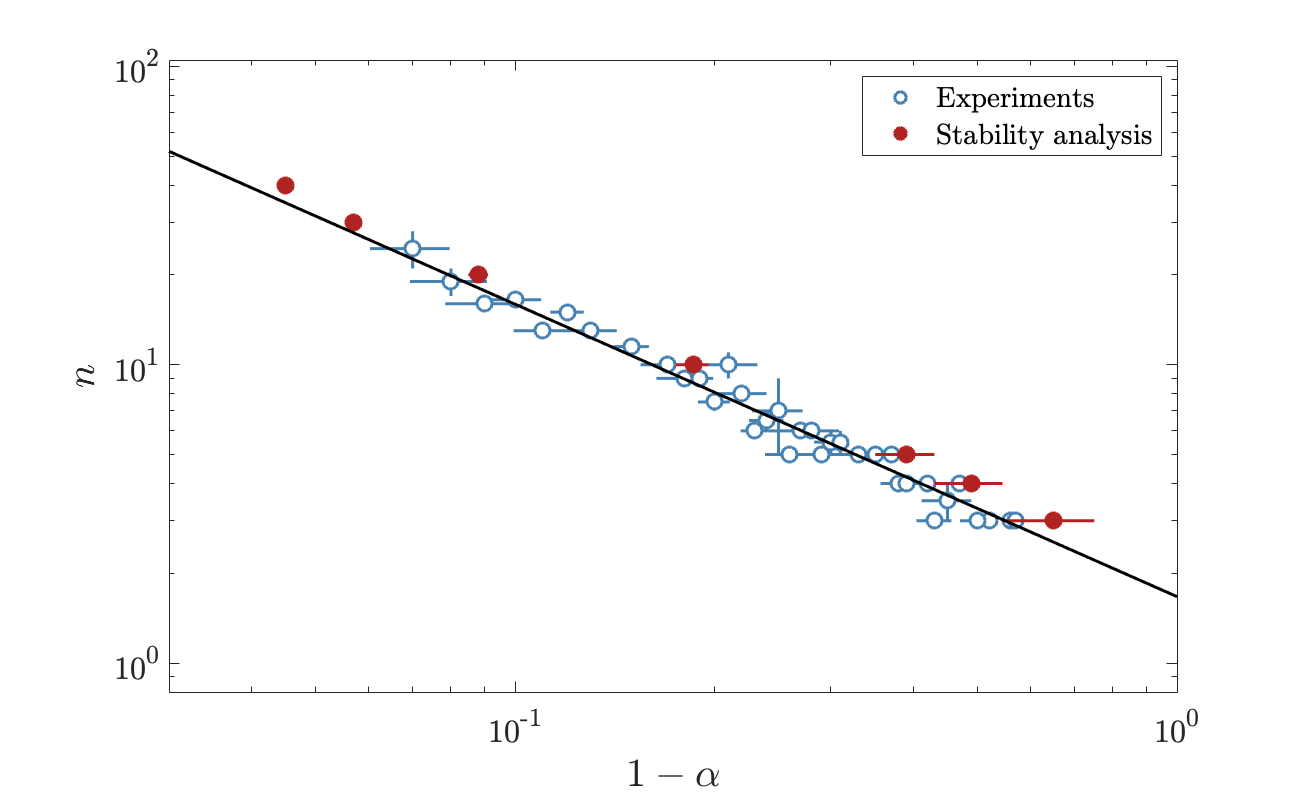}
\caption{Evolution of most unstable mode number as a function of $1-\alpha$. The symbols are computed using linear stability analysis and determining the range of $\alpha$ for which a mode is linearly the most unstable. A subset of experimental data is also shown where we represent the range of observed modes at threshold. Errors bars in $\alpha$ represent the uncertainty in the experimental aspect ratio. The solid line is the interpolation of the experimental data $n=1.67(1-\alpha)^{-0.98}$ in very good agreement with the scaling predicted in the $\alpha\rightarrow 1$ limit (\ref{sec_alpha1}) and with \cite{Sader_2019}.}
\label{fig_alphaN}
\end{center}
\end{figure}

It is interesting to look at the evolution of the most unstable mode as a function of $\alpha$ : we show in \ref{sec_alpha1} that the optimal wavenumber diverges as $ n \sim \left(1-\alpha\right)^{-1} $ in the limit $\alpha\rightarrow 1$. This is clearly confirmed both by linear stability analysis and experiments in figure \ref{fig_alphaN}. 
The interpolation of the experimental data gives $n=1.67(1-\alpha)^{-0.98}$ in very good agreement with \cite{Sader_2019}.
Note that this scaling, derived from energy minimization, can also be interpreted by a simple argument : the wavy pattern amplitude decreases from a maximal amplitude at the outer boundary to zero amplitude at the inner boundary. Thus, the pattern can extend radially at most on a length scale given by the width of the annulus. On the other hand, it is known that persistence length $L$ of a wavy pattern depends on the wavelength $\lambda$ and the amplitude $A$ \cite{Vandeparre_2011} : $L/\lambda = (A/h)^{1/2}$. Thus,
assuming that the wavelength of the pattern is of the same size that the width of the annulus gives $\lambda = \frac{2\pi b}{n} \sim (b-a)$, which upon inversion yields $n \sim \frac{1}{1-\alpha}$.

\begin{figure}
    \centering
    \begin{tabular}{cc}
 \includegraphics[scale=0.3]{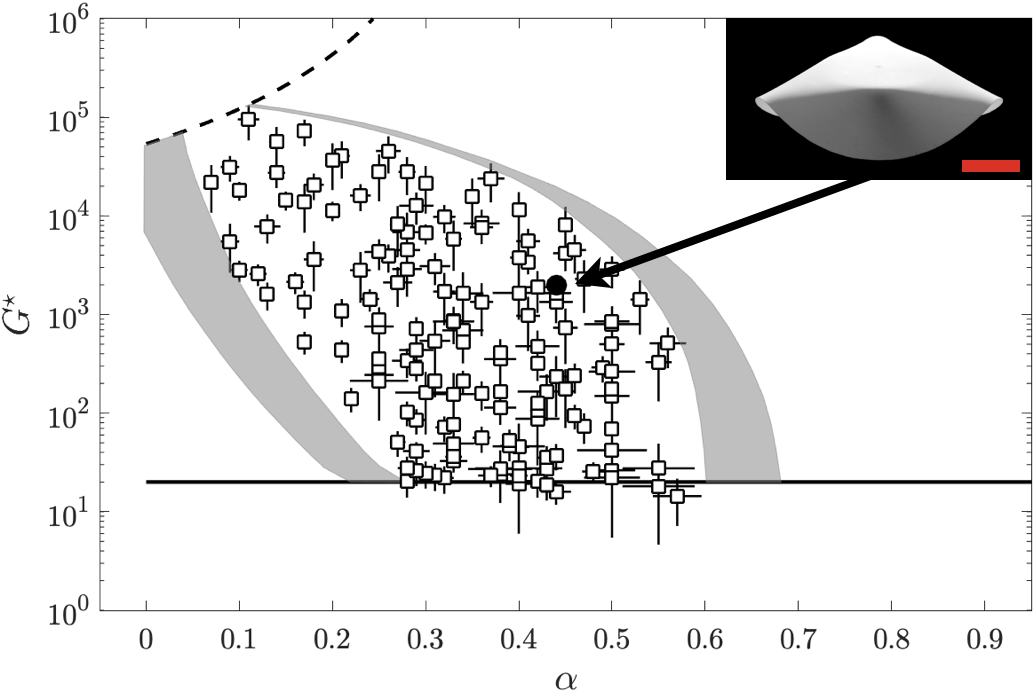} & 
 \includegraphics[scale=0.3]{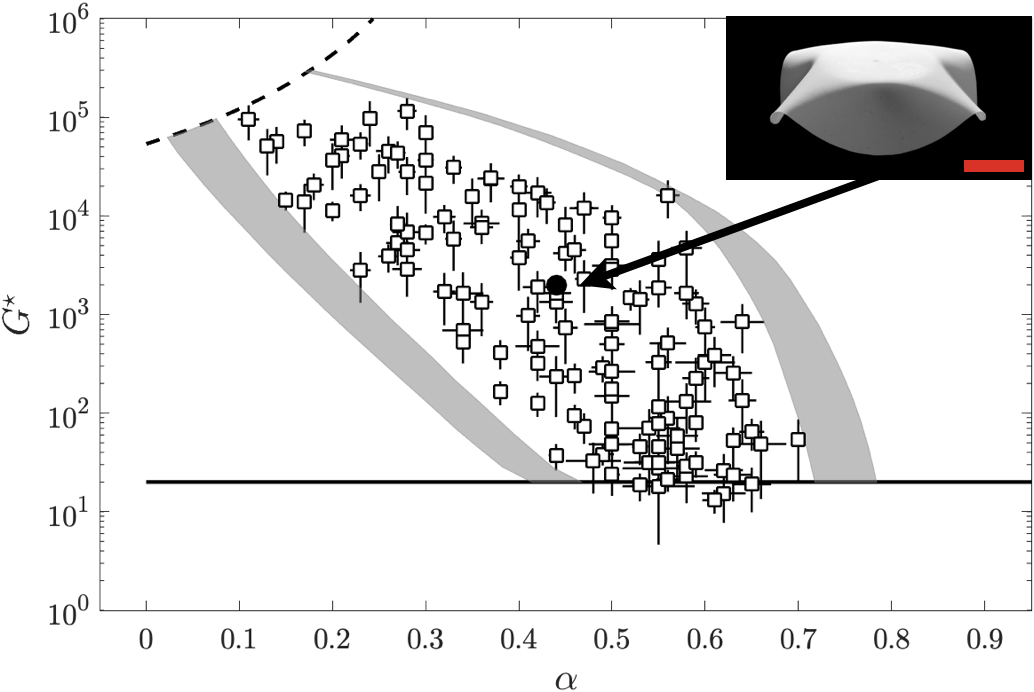} \\
 a & b\\
  \includegraphics[scale=0.3]{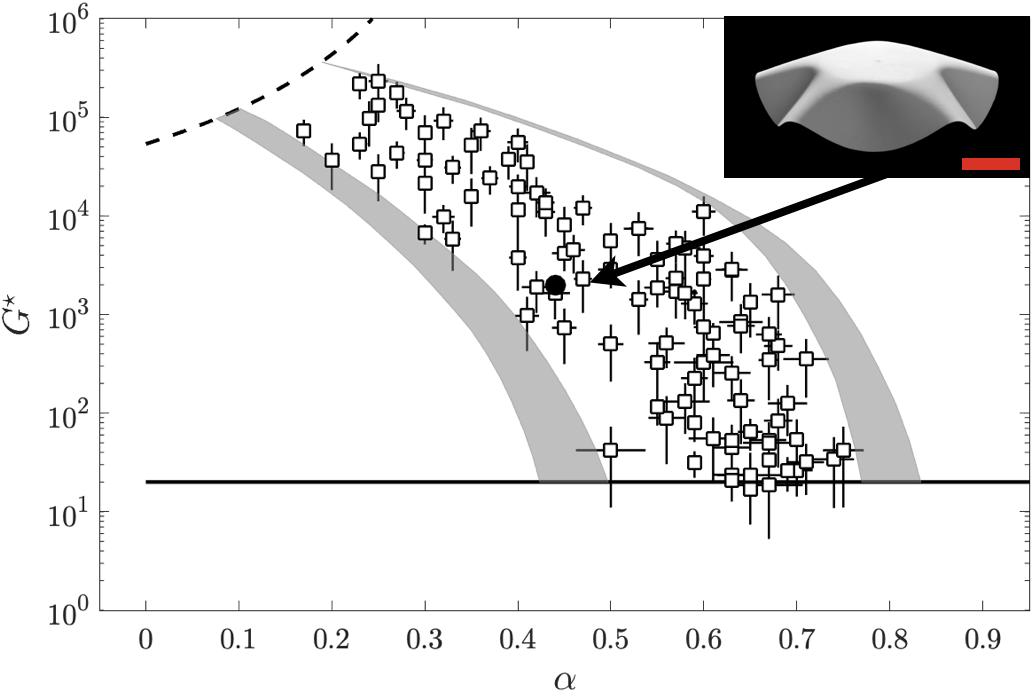} & 
   \includegraphics[scale=0.3]{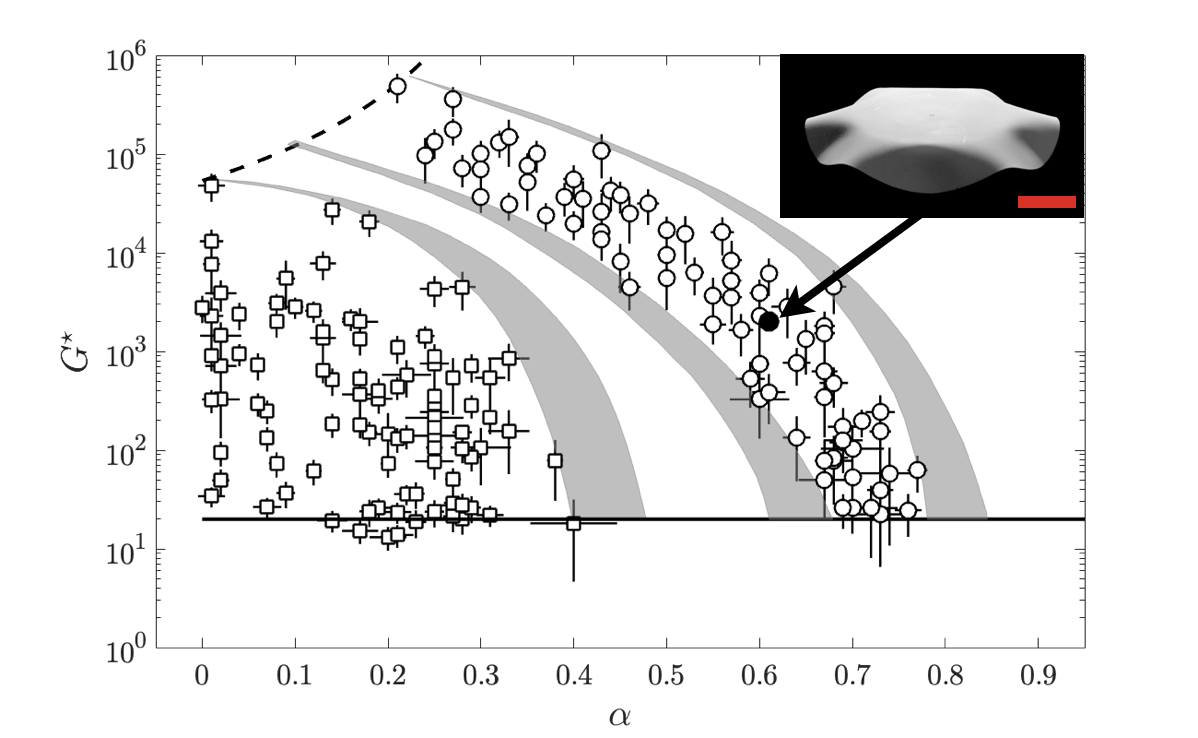} \\
  c & d\\
\end{tabular}
    \caption{Experimental diagram of the zone of existence of different modes (from $n=2$ to $n=6$). Pictures of experiments are also shown in inset, with the corresponding point highlighted in the diagram. Grey areas indicate the boundaries. a. mode $n=3$. b. mode $n=4$. c. mode $n=5$. d. mode $n=2$ (squares) and mode $n=6$ (circles). One can notice that zones of stability of modes overlap with each other, leading to multi-stability, as it is the case for the pictures of modes 3, 4 and 5 which are all obtained at $G^{\star}=2000$ and $\alpha=0.44$. The picture of mode 6 is obtained at $G^{\star}=2000$ and $\alpha=0.61$.  Scale bar is 20 mm.}
    \label{fig:ModesBoundaries}
\end{figure}

Finally, one can also notice that regions where axisymmetric state is linearly unstable to only one mode are the exceptions rather than the rule : for the most of the diagram, several modes can grow and interact nonlinearly. Thus, we analyze the modes selection experimentally by perturbing by hand the outer edge. For small aspect ratios and weak gravity, mode 2 (figure \ref{fig_photosXP_num} a) is the only mode observed: even if other modes are linearly unstable, they always destabilize to form a mode 2. For $\alpha \rightarrow 0$, increasing gravitational effects, a symmetric mode 2 can destabilize into an asymmetric one, consistent with \cite{Cerda_2004} predictions, but is often associated with self-contact, a regime we do not explore in this paper (see figure \ref{fig_photosXP_num} b). Increasing the aspect ratio, modes higher than $2$ can be observed. We construct the experimental zone of existence of different modes, from $n=2$ to $n=6$, as shown in figure \ref{fig:ModesBoundaries}. A given mode number can be observed at threshold (close to $G^\star=cst)$ over a finite range of aspect ratio. As $n$ increases, this range is both narrower and higher. Increasing $G^\star$ further from threshold, the range of $\alpha$ where a mode can be observed shifts toward lower values, until it hits the self-contact zone. Clearly, zones of existence of different modes overlap, which means that the system is multi-stable as illustrated for the point ($G^\star=2000, \alpha=0.44 $) where modes 3, 4 and 5 can be observed.

\section{Discussion and perspectives}

The deflection of a circular annulus by gravity leads to a wrinkling instability of the outer edge, where geometry plays a crucial role to determine both the threshold and the optimal wavenumber. We have shown that this influence can be caught in a modified dimensionless parameter $G^\star$ which express the ratio between normal load and bending forces. Using this parameter, the instability threshold is well described over the whole range of aspect ratio by a simple criterion $G^\star\approx 20$. To interpret this criteria, we develop a scaling analysis showing that wrinkling is associated with a critical deflection of the outer edge (which generates circumferential stresses). This deflection is governed by bending both in the $\alpha \rightarrow 0$ limit (where deflection are small enough for stretching effects to be negligible) and in the $\alpha \rightarrow 1$ limit (where radial displacement cancels the stretching energy to leading order in ($1-\alpha$), meaning that $\delta/h \sim G^\star$ over the whole range of $\alpha$ as shown in \ref{sec_alpha0} and \ref{sec_alpha1}.

\begin{figure}
    \centering
    \includegraphics[scale=0.25]{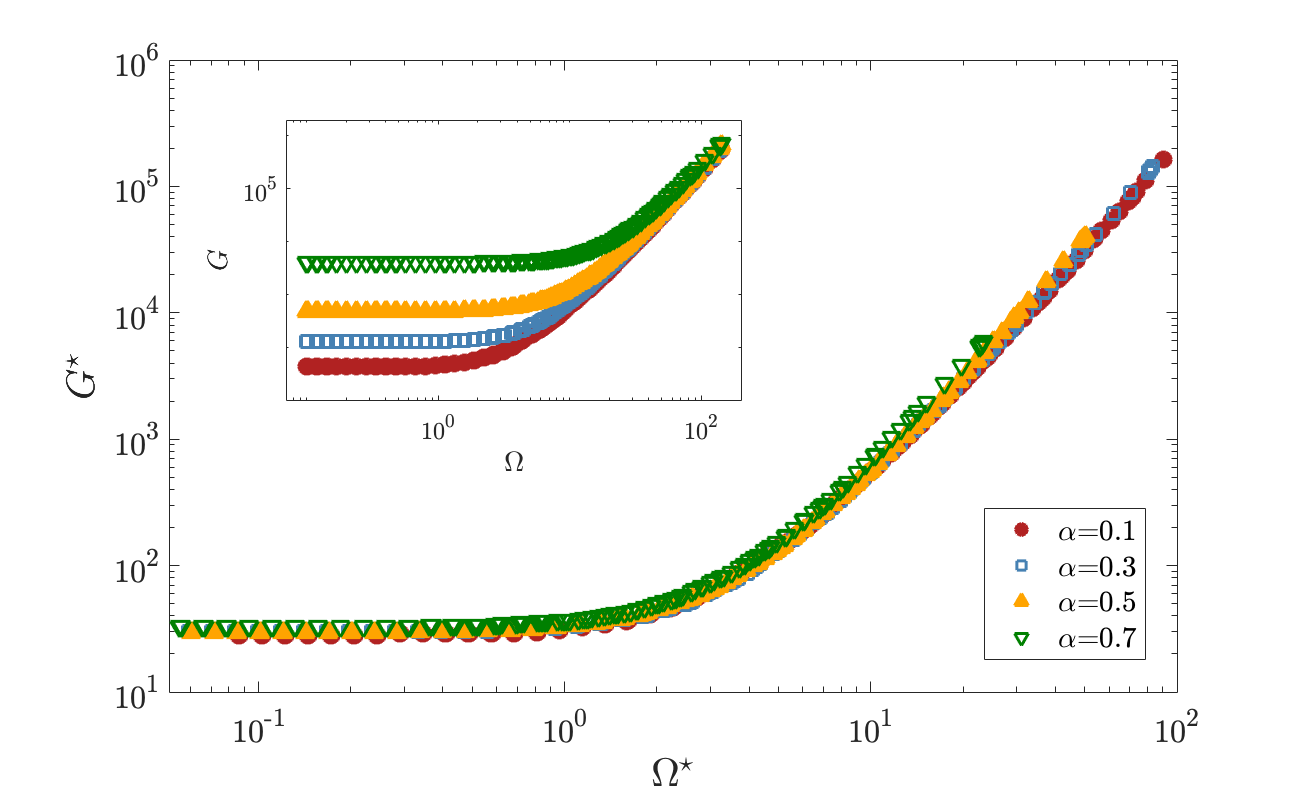}
    \caption{Critical curves $G,\Omega$ of \cite{Delapierre_2018} rescaled using the parameters $G^\star,\Omega^\star$ defined in the text. (Inset) original data of \cite{Delapierre_2018}}
    \label{fig:DelapierreRescaled}
\end{figure}

One possible attractive extension of this study would be to include other external forces, like centrifugal ones, as these forces are relevant in the deployment and stability of spacecraft structures \cite{Delapierre_2018}. These forces introduce another dimensionless parameter \cite{Chen_2011,Delapierre_2018} $\Omega = \sqrt{\frac{12(1-\nu^2)\rho h}{Eh^3}}b^2 \omega$. Introducing as in \ref{sec_stabilityAnalysis} the distinction between radial length scale $b$ and width of the annulus $b(1-\alpha)$ for derivation, one is led to the following modifications: $\Omega^\star = \Omega (1-\alpha)^{3/2}$ in the $\alpha \rightarrow 0$ limit and $\Omega^\star = \Omega (1-\alpha)^{2}$ in the $\alpha \rightarrow 1$ limit. We test the former scaling by using the results of \cite{Delapierre_2018} where the critical curves $G(\Omega)$ obtained by theoretical analysis were reported for four different values of $\alpha$ : extracting the data from figure 12a of \cite{Delapierre_2018} and replotting it (figure \ref{fig:DelapierreRescaled}) in terms of $G^\star$ and $\Omega^\star$ indicates that the scaling collapses the data onto a single curve. Even if it is mathematically valid only in the $\alpha \rightarrow 0$ limit, it seems to hold for values of $\alpha$ as high as $0.7$. It would be interesting to test this scaling for more values of $\alpha$ and test it experimentally , especially close to $\alpha \approx 1$ to see whether or not there is an universal critical curve valid for all aspect ratio, a fact that might be useful for the design of such structures. 

From a more fundamental point of view, 
the analysis developed here may also extend to the classical Lam\'e setup, where a similar blowup of eigenvalues as a function of aspect ratio is observed \cite{Coman_2007}. Including the aspect ratio into the control parameter may help to delineate the respective influence of geometry and physics.

\section*{Acknowledgements}
Centre de Calcul Intensif d'Aix-Marseille is acknowledged for granting access to its high performance computing resources.


\appendix
\section{The $\alpha\rightarrow 0$ limit}
\label{sec_alpha0}


\subsection*{Unwrinkled state}

In the limit of a point-like support, there is only one length scale $b=l$. 
The radial displacement is determined by minimizing the total energy with respect to $U$, which is equivalent to minimizing the stretching energy with respect to $U$ because at first order the radial displacement $U$ appears only in the stretching energy. Thus, differentiating 
equation (\ref{eq_estretch0}) gives :
$$U \sim -\frac{\delta^2}{4 b} $$

The bending energy of an unwrinkled sheet can be estimated as $\mathcal{E}^{b}\sim 2 S Eh^3  \left(  \frac{\delta}{b^2}\right)^2$.
Thus, the typical vertical deflection solves the minimization of the total (stretching + bending + gravitational) energy, which reads
$$ \frac{Eh\delta^3}{2b^4}+\frac{E\delta h^3}{3b^4}-\rho g h=0$$
which gives two different regimes. For $\frac{\delta}{h}\ll 1$, bending is dominant and $\frac{\delta}{h} \sim G^\star$, while for $\frac{\delta}{h}\gg 1$, stretching is dominant and $\frac{\delta}{h} \sim G^{\star \frac{1}{3}}$

\subsection*{Wrinkling threshold}



We now estimate the threshold for wrinkling by inserting a non zero amplitude modulation : $z(r,\theta) = \delta(1+A\cos n \theta)$

At dominant order, the modulation appears in $\epsilon_{\theta\theta} \sim \frac{U}{b}+\frac{\delta^2 A^2n^2}{2b^2}\sin(n \theta)^2$ and in $\kappa_{\theta\theta} \sim \frac{\delta}{b^2} - \frac{\delta An^2}{b^2}\cos(n \theta)$.  With this in mind, and recalling that $U\sim -\frac{\delta
^2}{b}$, we clearly see the physical mechanism responsible for wrinkling : elastic energy due to the orthoradial stretching $$\mathcal{E}^{s}_{\theta\theta}\sim E h \int_S \epsilon_{\theta\theta}^2 \D S \sim Eh \int_S \left(-\frac{\delta
^2}{b^2} +\frac{\delta ^2A^2n^2}{2b^2}\sin(n\theta)^2 \right)^2 \D S$$
can be diminished by allowing wrinkles to grow. However, this is balanced by a cost in bending energy
$$\mathcal{E}^{b}_{\theta\theta} \sim \frac{Eh^3}{12} \int_S  \left(\frac{\delta}{b^2} - \frac{\delta An^2}{b^2}\cos(n \theta)\right)^2 \D S$$
Expanding and then integrating over $\theta$, simple algebra permits to obtain the $A$-dependent terms of the energy as 
$$\mathcal{E}_A \sim \left[ \frac{3A^4\delta^4n^4}{16b^4}- \frac{A^2 \delta^4n^2}{4b^4}+ \frac{h^2}{12} \frac{A^2\delta^2n^4}{b^4} \right] E hS $$

Thus, amplitude can be found by minimization of the energy with respect to $A$ which leads to 
\begin{equation}
    A=0 \quad \text{or}\quad  A \sim \sqrt{\frac{1}{n^2}-\frac{1}{3}\left(\frac{h}{\delta}\right)^2}
\end{equation}

Positivity of the square root requires that $\frac{\delta}{h} > \frac{n}{\sqrt{3}}$, meaning that there is a threshold value of the deflection below which axisymmetric state ($A=0$) is stable. 

\section{The $\alpha \rightarrow 1$ limit}
\label{sec_alpha1}

\subsection*{Unwrinkled state}
In the limit of an elastic disk slightly larger than its support, the orthoradial stretching $\epsilon_{\theta\theta} \sim \frac{U}{b}$ is subdominant compared to radial stretching $\epsilon_{rr} \sim \frac{U}{(1-\alpha)b} + \frac{\delta^2}{2 (1-\alpha)^2b^2}$, which is the dominant term in the stretching energy. Thus, radial displacement is found by minimization of radial stretching 
\begin{equation}
U\sim -\frac{\delta^2}{2(1-\alpha)b}
\label{eq_Ualpha1}
\end{equation}
and cancels to leading order the stretching energy. Deflection under gravity is then governed by bending only and leads to $\frac{\delta}{h}\sim G^\star$

\subsection*{Wrinkling threshold}
Inserting the expression for radial displacement (\ref{eq_Ualpha1}) in the stretching energy, one can compute the next order of the stretching energy in powers of $(1-\alpha)$. Minimizing the total energy with respect to amplitude and mode number yields the following scalings :
\begin{equation}
\begin{split}
    n &\sim (1-\alpha)^{-1} \\
    A & \sim \sqrt{3\sqrt{2}(1-\alpha)-\left(\frac{h}{\delta}\right)^2}
\end{split}    
\label{eq_scalings}
\end{equation}

Again, for a fixed aspect ratio, there is a critical deflection for which wrinkling is energetically favorable. The most favorable mode number is diverging as a function of $1-\alpha$, which is clearly confirmed by figure \ref{fig_alphaN} where we represent for each mode the range of $\alpha$ where it is the most dangerous. There is a very good agreement between this scaling and experimental data, as well as values coming from the stability analysis (which keeps all the complexity of the F\"oppl-von Karman equations).
Note that the scaling (\ref{eq_scalings}) predicts that $\frac{\delta}{h} \sim (1-\alpha)^{-\frac{1}{2}}$, which means that the critical $G^\star$ should be a weakly diverging function of $(1-\alpha)$. This is consistent with what we observe in linear stability analysis and full numerical simulations.

\section*{References}

\bibliography{biblio}

\end{document}